\documentclass[%
 rsi,
 amsmath,amssymb,
 reprint,%
]{revtex4-1}

\usepackage{graphicx}% Include figure files
\usepackage{dcolumn}% Align table columns on decimal point
\usepackage{bm}% bold math
\usepackage[utf8]{inputenc}
\usepackage[T1]{fontenc}
\usepackage{mathptmx}
\usepackage{amsmath}
\usepackage{etoolbox}
\usepackage{comment}
\usepackage[dvipsnames]{xcolor}
\usepackage{ amssymb }
\usepackage[caption=false]{subfig}
\usepackage[title]{appendix}
\usepackage{enumerate}% http://ctan.org/pkg/enumerate
\usepackage{float}
\usepackage{enumitem}
\usepackage{physics}
\usepackage{caption}
\usepackage{subcaption}
\makeatletter
\def\@email#1#2{%
 \endgroup
 \patchcmd{\titleblock@produce}
  {\frontmatter@RRAPformat}
  {\frontmatter@RRAPformat{\produce@RRAP{*#1\href{mailto:#2}{#2}}}\frontmatter@RRAPformat}
  {}{}
}%
\makeatother
\begin{document}
\preprint{AIP/123-QED}

\title{XCOM: Full Mesh Network Synchronization and Low-Latency Communication for QICK (Quantum Instrumentation Control Kit) }

% Force line breaks with \\

\author{Diego Martin}
\affiliation{Fermi National Accelerator Laboratory, Batavia IL, 60510, United States}
\author{Luis H. Arnaldi}
\affiliation{Fermi National Accelerator Laboratory, Batavia IL, 60510, United States}
\author{Kenneth Treptow}
\affiliation{Fermi National Accelerator Laboratory, Batavia IL, 60510, United States}%Lines break automatically or can be forced with \\
\author{Neal Wilcer}
\affiliation{Fermi National Accelerator Laboratory, Batavia IL, 60510, United States}%Lines break automatically or can be forced with \\
\author{Sho Uemura}
\affiliation{Fermi National Accelerator Laboratory, Batavia IL, 60510, United States}%Lines break automatically or can be forced with \\
\author{Sara Sussman}
\affiliation{Fermi National Accelerator Laboratory, Batavia IL, 60510, United States}%Lines break automatically or can be forced with % 
\author{David I Schuster}
\affiliation{Department of Physics and Applied Physics, Stanford University, Stanford CA, 94305}%Lines break automatically or can be 
\author{Gustavo Cancelo*}
\affiliation{Fermi National Accelerator Laboratory, Batavia IL, 60510, United States}
\email{cancelo@fnal.gov.}

\date{\today}% It is always \today, today,

\begin{abstract}
Quantum computing experiments and testbeds with large qubit counts have until recently been a privilege afforded only to large companies ~\cite{google2024,AbuGhanem_2025} or quantum technologies where scaling to hundreds or thousands of qubits does not require a substantial increase in quantum control hardware (neutral atoms~\cite{kornja,zhang2025}, trapped ions~\cite{Guo_2024}, or spin defects~\cite{jdzq-jbfz}). Superconducting and spin qubit testbeds critically depend on scaling their control systems beyond what a single electronics board can provide. Multi-board control systems combining RF, fast DC control, bias, and readout require precise synchronization and communication across many hardware and firmware components. To address this, we present XCOM, a network that synchronizes QICK boards \cite{Stefanazzi2022,Ding2024} and the absolute clocks governing quantum program execution to within 100\,ps, free of drift and loss of lock. XCOM also provides deterministic, all-to-all simultaneous data communication with latency below 185\,ns. Like QICK itself, XCOM is compatible with a broad range of qubit technologies and is designed to scale to large systems.
\end{abstract}

% \begin{abstract}
% Until now quantum computer experiments and testbeds with a large number of qubits have been a privilege of big companies (E.g.~\cite{google2024},~\cite{AbuGhanem_2025}) or quantum technologies where scaling up to hundreds or thousands of qubits does not require a big jump in quantum control equipment (E.g. AMOs~\cite{kornja}~\cite{zhang2025}, trapped-ions ~\cite{Guo_2024}, NV-centers~\cite{jdzq-jbfz}). Superconducting and spin qubit testbeds critically depend on scaling up their control systems beyond what a single electronic board can provide. A multi-board approach combining RF of fast DC control, bias, and readout requires synchronizing and communicating many pieces of hardware and firmware. The XCOM network synchronizes and provides a deterministic, low latency communication path among QICK boxes~\cite{Stefanazzi2022}~\cite{Ding2024}. XCOM synchronizes boards and the absolute clock that times quantum program execution to less than 100ps. XCOM is free of drift and loss of reference. XCOM allows all-to-all simultaneous communication of data with deterministic latency below 185ns. XCOM, as QICK, can be used for large number of qubit technologies and scales.
% \end{abstract}

\maketitle

\section{Introduction}\label{Intro}

Among today's successful qubit technologies, superconducting qubits are notable for requiring a very high hardware I/O count~\cite{Castelvecchi}. Silicon spin qubits and quantum dots are also very demanding in the number of fast DC bias outputs required to manipulate spins and perform readout~\cite{Weinstein2023,George_2024}. Quantum hardware continues to evolve at a rapid pace. Control systems need to be flexible, while also providing high performance and complex functionality. FPGAs continue to be at the center of modern control systems; in particular, RF-FPGAs help integrate analog RF acquisition and generation of signals up to 10\,GHz with high-speed parallel logic, minimizing hardware footprint and power consumption~\cite{AMDrfsoc}. In the future, ASICs will take over some of the functions of 
today's room-temperature electronics, but they still need to 
overcome several challenges. Modern RF-FPGAs such as the AMD 
RFSoC integrate high-speed DACs, ADCs, and programmable logic 
in a single device, a level of complexity and reconfigurability 
that cryo-CMOS ASICs have yet to match. We expect that for the next 5 to 10 years, room-temperature FPGAs and discrete component electronics will continue to drive the ecosystem~\cite{AMDversalrf}.

In order to minimize physical channel count, QICK has achieved 8- and 16-channel frequency multiplexing. This is successfully used for multi-qubit readout~\cite{Ding2024}. Multiplexing of qubit control remains a qubit device challenge. Today, most proprietary, commercial, and open-source control systems map from a few up to 16 signal generator outputs (plus inputs) to a single FPGA. It is clear that larger qubit systems require control hardware expansion, hosting multiple FPGAs, DACs, ADCs, bias circuits, I/O, and computer interface ports (e.g., Ethernet, USB, etc.). Furthermore, most systems requiring multi-piece hardware platforms also require the hardware pieces to be synchronized and to support communication. In this publication, we describe XCOM, a synchronization and low-latency message exchange bus with a mesh network topology.
Multi-module hardware communication is available in most modern quantum control systems. For instance, Xu et al.\ at LBNL have reported a comprehensive list of control systems that support synchronization and communication at some level~\cite{xu2025multifpga}. In particular, the QubiC project~\cite{xu2021} and the Manarat 
project~\cite{silva2025manarat} use the AMD RFSoC FPGAs and ZCU216 evaluation boards that are also the hardware foundation of QICK~\cite{rfsocdatasheet}.

\begin{figure}[h!]
     \centering
     \includegraphics[width=\columnwidth]{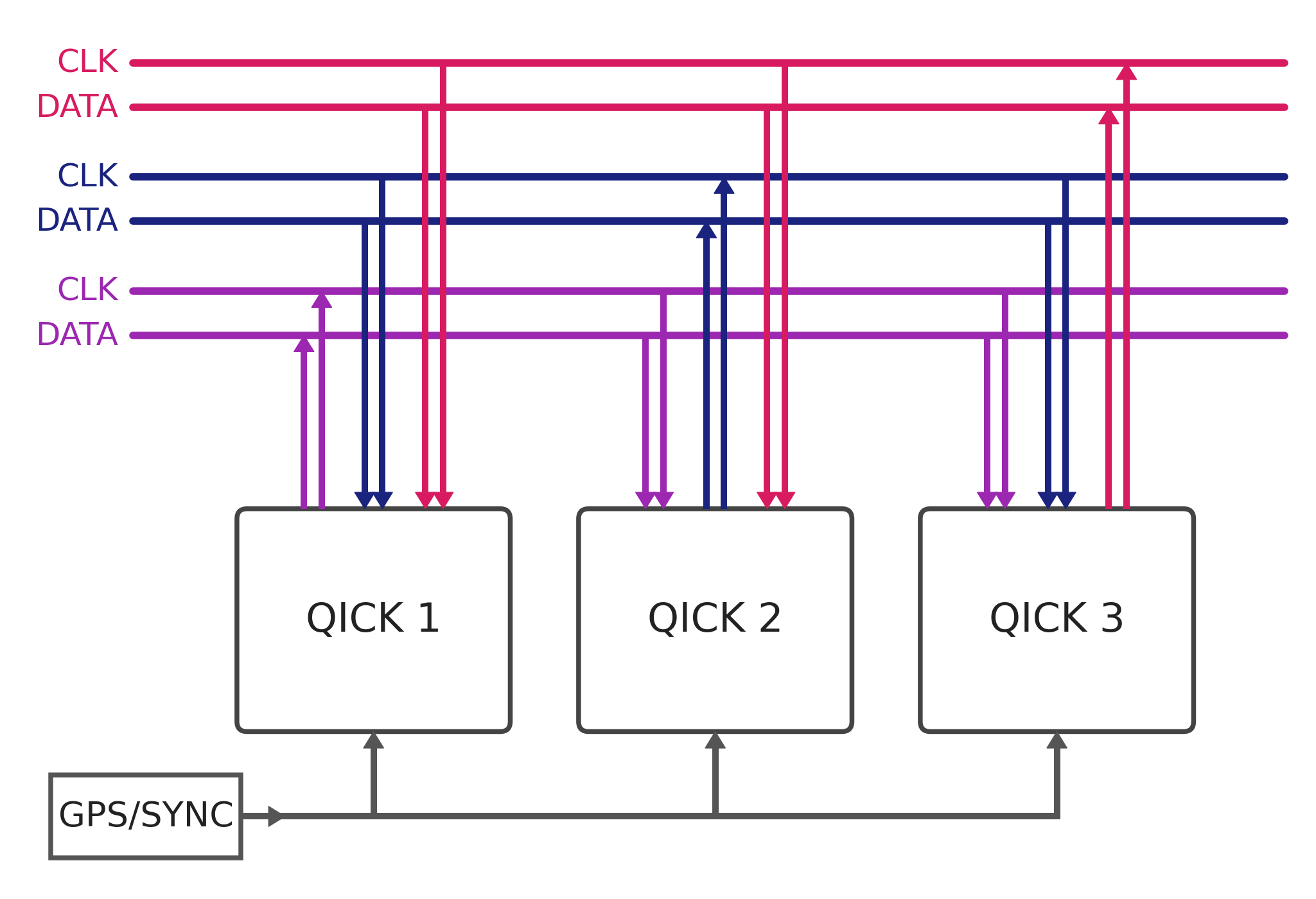}
     \captionsetup{justification=raggedright}
      \caption{XCOM block diagram. Each QICK board connects a \texttt{Tx$_{\texttt{data/clk}}$} output to a pair of LVDS lines. Three \texttt{Tx$_{\texttt{data/clk}}$} channels are shown in purple, blue, and red. All boards have \texttt{Rx$_{\texttt{data/clk}}$} inputs to listen to all \texttt{Tx$_{\texttt{data/clk}}$} channels. A shared GPS/SYNC reference (bottom) provides the external clock common to all boards.}
    \label{fig:XCOM_BD}
\end{figure}

 \section{Architecture Overview}\label{sec:XCOMdescription}

Each QICK board runs its own timed-processor (tProc), which maintains a 48-bit absolute clock counter used to time all quantum program execution (i.e. the experiment clock domain). While all boards share the same clock frequency, their absolute counters are not inherently aligned. This misalignment has direct consequences for multi-board experiments: for instance, a \texttt{pulse} command issued simultaneously on two different boards will not produce temporally aligned output pulses unless the absolute clocks on both boards have been synchronized. XCOM is designed to solve this problem.
 
The XCOM architecture is a full mesh network as shown in Figure~\ref{fig:XCOM_BD}. There are as many data channel/clock pairs as there are QICK boards connected to the network. Each board is the sole speaker on its own channel, while all boards listen on all channels. This parallel architecture allows point-to-point and broadcast messages to occur concurrently on different channels. Received messages are stored in memory and handled by the XCOM peripheral IP firmware block as described in Section~\ref{sec:XCOM_sync_sec}. Prior to synchronization, each board is assigned a network ID number, which can be set by software or auto-assigned by the XCOM.
 
To synchronize the network, one board is designated the master. The master is allowed to broadcast reset, start, and stop commands to the absolute clocks of all boards, aligning every tProc to the same counter value and enabling precise cross-board pulse timing. Since all XCOM interfaces and firmware IPs are identical, any board can be assigned the master role in software. Board frequency synchronization is based on an external reference source (e.g., a rubidium clock), and a single broadcast from the master is sufficient to synchronize all absolute time counters across the network simultaneously. The board synchronization is required to happen a single time, when the XCOM network is configured or when a new node is added to the network.

\begin{figure}[h!]
\begin{flushleft}
(a)\\[2pt]
\includegraphics[width=\columnwidth]{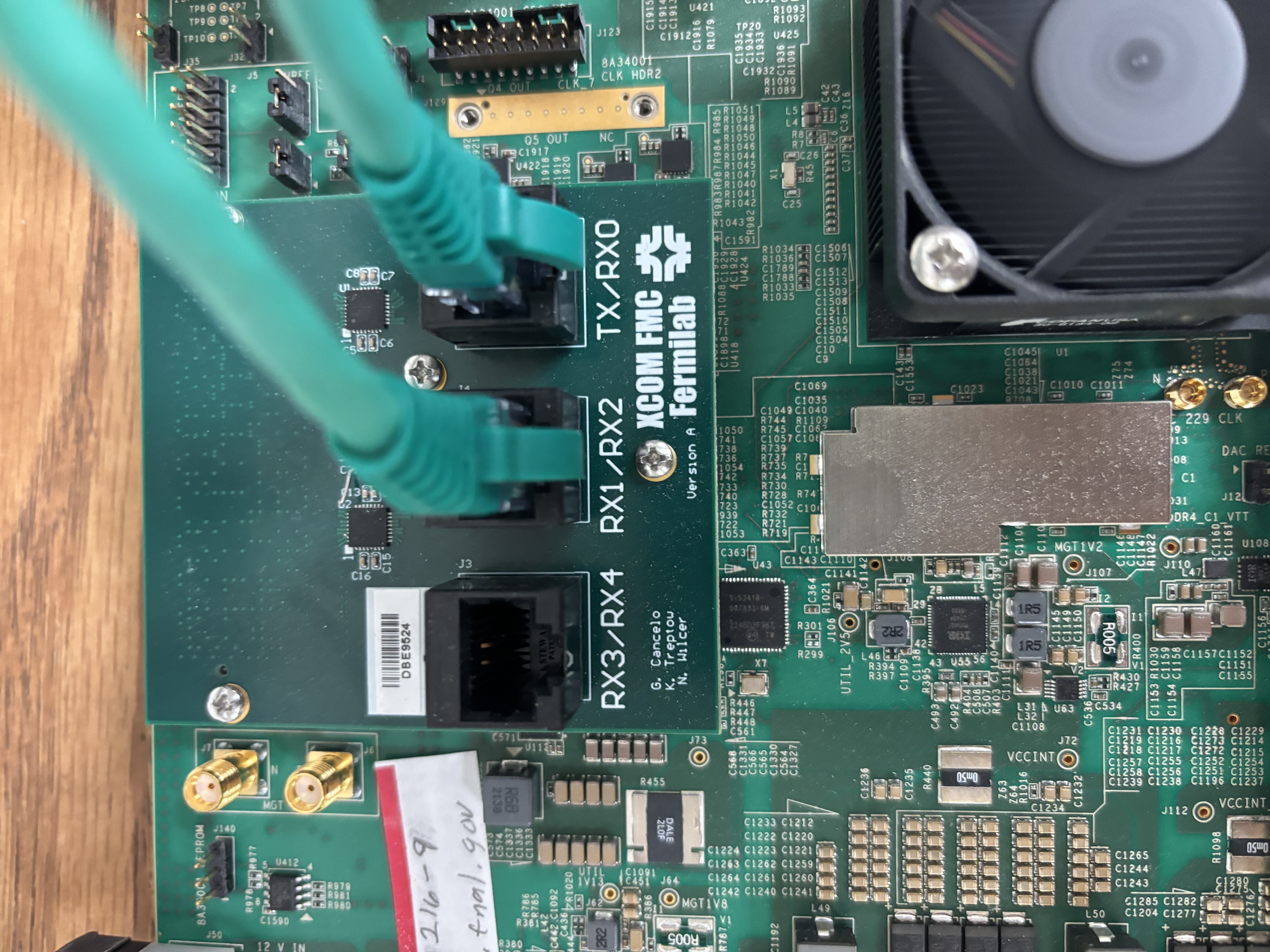}
(b)\\[2pt]
\includegraphics[width=\columnwidth]{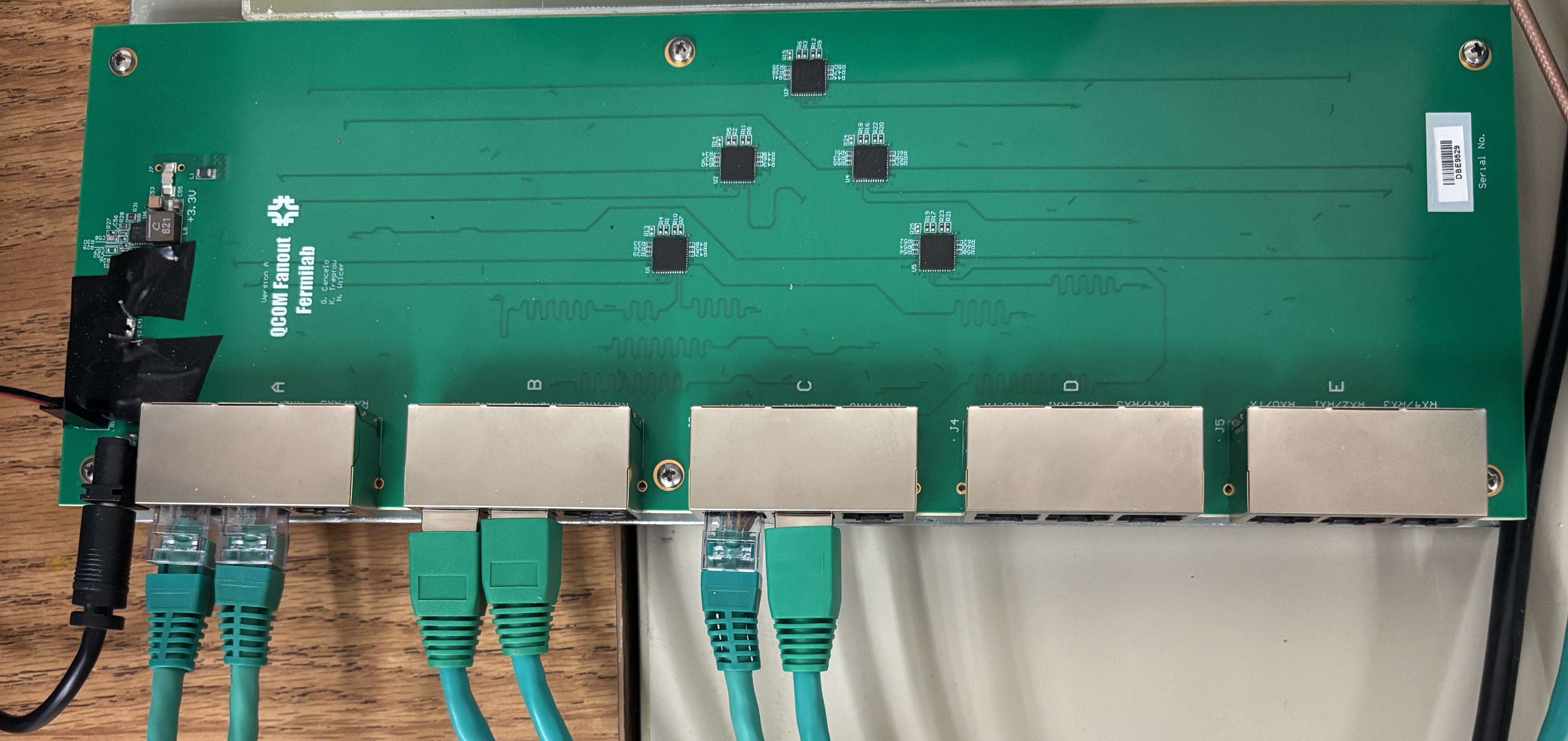}
\captionsetup{justification=raggedright}
\caption{XCOM hardware. (a) Transceiver board mounted on the RFSoC FMC
connector, providing one \texttt{Tx$_{\texttt{data/clk}}$} driver and five \texttt{Rx$_{\texttt{data/clk}}$}
receivers via three RJ45 connectors. One such board is required per QICK
board in the network. (b) Standalone fanout hub, which distributes each
incoming \texttt{Tx$_{\texttt{data/clk}}$} pair to five \texttt{Rx$_{\texttt{data/clk}}$} output pairs, one
per board in the network.}
\label{fig:XCOMhw}
\end{flushleft}
\end{figure}

\section{Hardware Implementation}\label{sec:XCOM_hardware}

The XCOM hardware consists of two components: a small transceiver board 
plugged into the FMC connector of the ZCU216 board (Figure~\ref{fig:XCOMhw}a), 
and a single external hub to fan out \texttt{Tx} drivers from each board to 
\texttt{Rx} receivers in all boards (Figure~\ref{fig:XCOMhw}b). The current 
hardware prototype supports up to five RFSoC boards, while the XCOM 
firmware IP already supports up to 15 boards and can be extended further; 
a larger hardware prototype is straightforward to build.

The XCOM maps a single \texttt{Tx$_{\texttt{data/clk}}$} from each board to 5 \texttt{Rx$_{\texttt{data/clk}}$}
in all boards. To minimize FPGA logic and data latency, we implement a 
separate clock per \texttt{Tx} driver. We choose this option over the one that 
embeds clock and data on a single LVDS line, because the latter requires 
extracting and synchronizing the transmitter clock to the receiver clock, 
adding more logic and latency.

XCOM uses LVDS high performance (HP) I/O from the FPGA routed to the FMC connector. The small XCOM \texttt{Tx/Rx} board plugs onto the FMC and uses the TI DS90LV804 LVDS
quad-drivers capable of driving 60mA at 800Mb/s to bring the
HP I/Os to the three RJ45 female connectors as shown in Figure~\ref{fig:XCOMhw}a. 
Each connector carries 
2 LVDS data channels and 2 LVDS clock pairs (e.g., \texttt{DATA$_{\texttt{0}}$}, \texttt{CLK$_{\texttt{0}}$} pair; 
\texttt{DATA$_{\texttt{1}}$}, \texttt{CLK$_{\texttt{1}}$} pair). The LVDS \texttt{Tx$_{\texttt{data/clk}}$} pairs are input to the 
XCOM hub, which uses TI LMK1D2108 buffers to fan out each pair to output 
\texttt{Rx$_{\texttt{data/clk}}$} signals routed back to the FMC board on each QICK board as shown in Figure~\ref{fig:XCOMhw}b.

\begin{figure}[h!]
\centering
\includegraphics[width=\columnwidth]{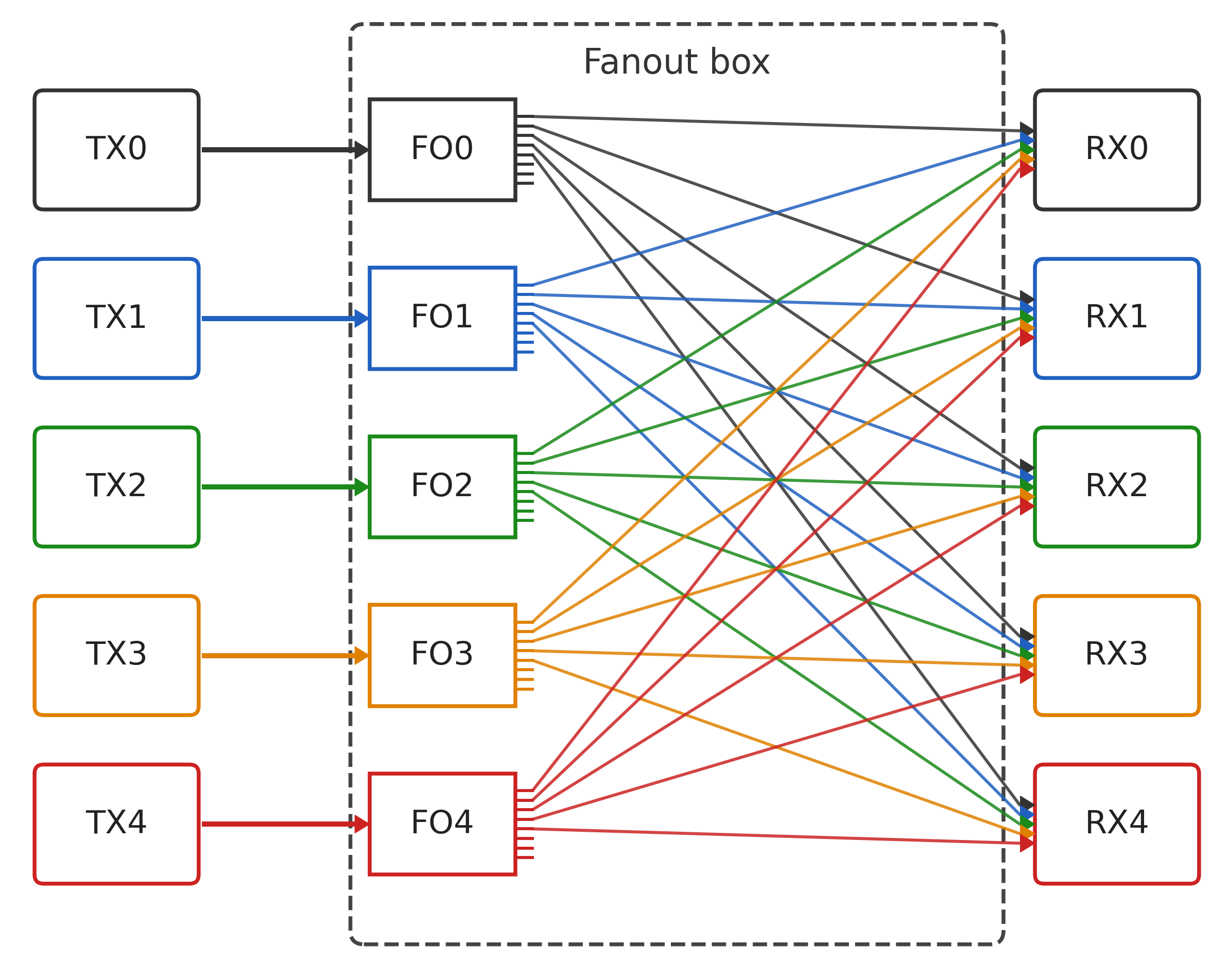}
\captionsetup{justification=raggedright}
\caption{Block diagram of the XCOM fanout hub. Each \texttt{Tx$_{\texttt{i}}$} 
($\texttt{i} = 0, \ldots, 4$) originates from a different RFSoC board and is 
copied five times by fanout buffer \texttt{FO$_{\texttt{i}}$}. Each output is routed to 
one \texttt{Rx$_{\texttt{i}}$} on a board in the network. \texttt{Tx$_{\texttt{i}}$} and \texttt{Rx$_{\texttt{i}}$} blocks sharing 
the same color reside on the same RFSoC FMC transceiver board 
(Figure~\ref{fig:XCOMhw}a).}\label{fig:fanout}
\end{figure}

Figure~\ref{fig:fanout} shows the block diagram of the XCOM fanout hub. 
Each \texttt{Tx$_{\texttt{i}}$} ($\texttt{i} = 0, \ldots, 4$) originates from a different RFSoC board 
and enters the hub, where it is copied by fanout buffer \texttt{FO$_{\texttt{i}}$} into five 
outputs. Each fanned-out output is routed to one \texttt{Rx$_{\texttt{i}}$} on each board. 
Note that \texttt{Tx$_{\texttt{i}}$} and \texttt{Rx$_{\texttt{i}}$} blocks sharing the same color reside on the 
same RFSoC FMC transceiver board (Figure~\ref{fig:XCOMhw}a).

The FPGA HP I/O LVDS supports up to 312.9\,MHz. For this prototype, we 
have initially set the clock to 100\,MHz and used both clock edges to 
sample data, achieving 200\,Mb/s. The clock speed will be pushed to 
300\,MHz to reduce the current 32-bit word latency from 186\,ns to 62\,ns.

\section{Synchronization Protocol}\label{sec:XCOM_sync_sec}

\begin{figure}[h!]
\centering
\includegraphics[width=\columnwidth]{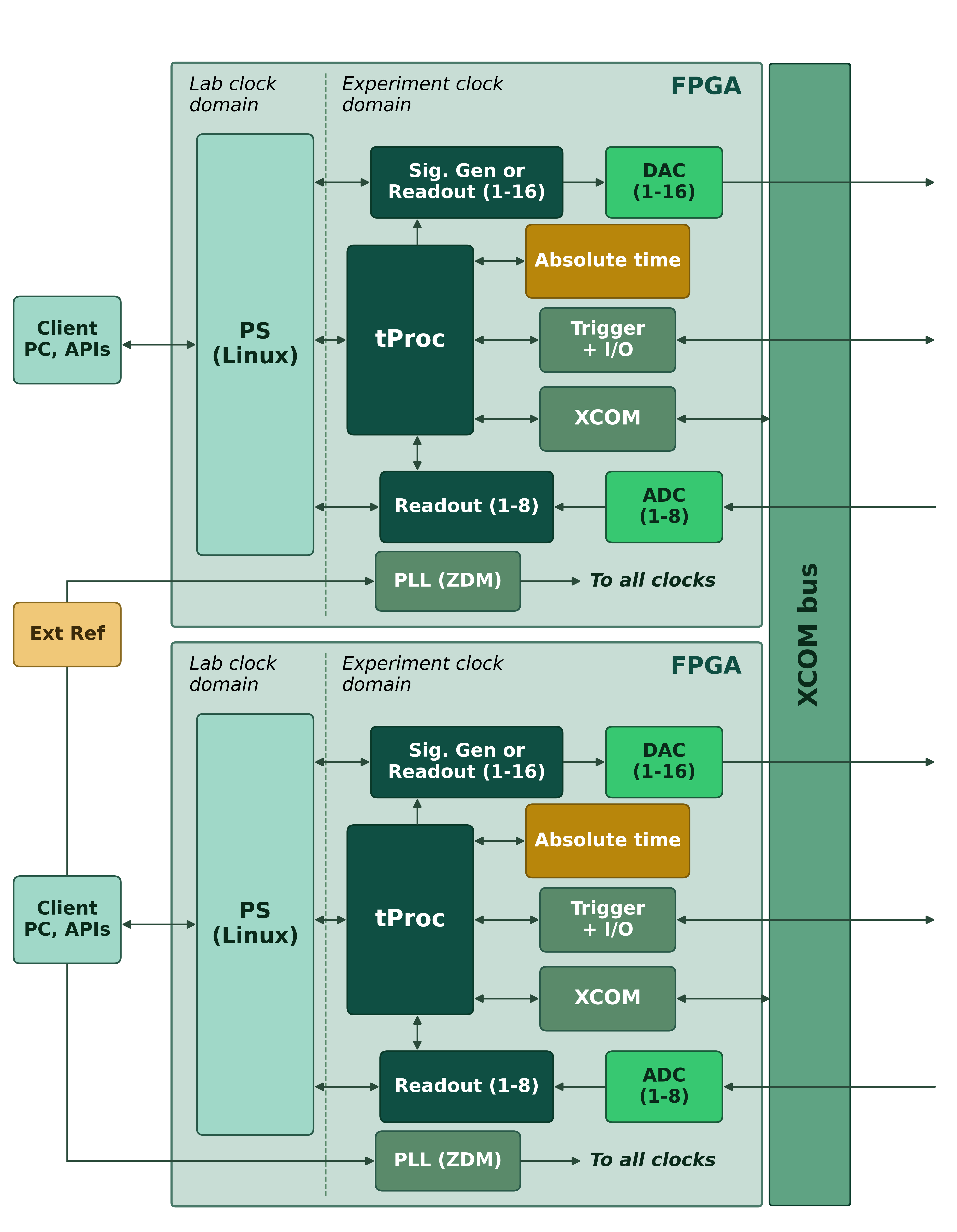}
\captionsetup{justification=raggedright}
\caption{QICK firmware block diagram. Two QICK firmware FPGA blocks (light green) 
are shown, each connected to a shared external stable reference and to the 
XCOM parallel bus. Each firmware block contains up to 16 Signal Generators 
(SG, with up to 16$\times$ frequency multiplexing), up to 8 Readouts (RD, 
with up to 8$\times$ frequency multiplexing), 16 digital I/O channels, and 
peripherals including the XCOM. Each block has a dedicated tProc and 
absolute time clock. The DACs and ADCs are internal to the RFSoC FPGA and 
communicate with the logic blocks at up to 10\,GS/s and 2.5\,GS/s, 
respectively. All blocks operate in the experiment clock domain. The QICK 
firmware communicates with the FPGA's built-in quad-core ARM processor (PS) 
and through it to the client PC.}\label{fig:XCOM_sync}
\end{figure}

\begin{figure*}
\centering
\includegraphics[width=0.97\textwidth]{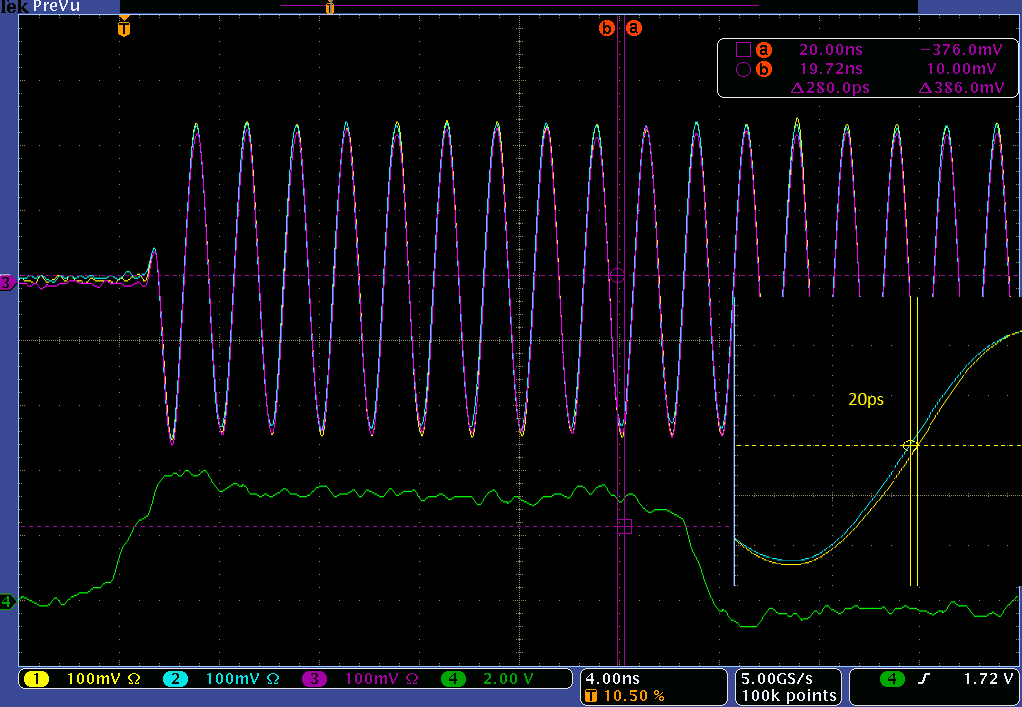}
\captionsetup{justification=raggedright}
\caption{Oscilloscope traces of three RF waveforms from three different 
QICK boards and one PMOD digital I/O signal, all triggered simultaneously 
via matched-length cables. The PMOD drivers are bandwidth-limited, showing 
a 1.5\,ns rise time. The inset shows a timing skew of only 20\,ps between 
the three RF channels, demonstrating sub-100\,ps cross-board 
synchronization. Achieving this level of synchronization requires PLL 
nested zero-delay mode lock on each board, multi-tile synchronization 
(MTS) across DAC and ADC tiles within each board, and absolute clock 
alignment across all boards via XCOM.}\label{fig:scope_RF}
\end{figure*}

The QICK main firmware components, as shown in Figure~\ref{fig:XCOM_sync}, are the Signal Generators, the Readouts, the tProc, and peripherals such as digital I/O and the XCOM.
The Signal Generators (SG) and Readouts (RD) have been described in~\cite{Ding2024} and~\cite{Stefanazzi2022}. SGs and RDs support multi-tone multiplexing of up to 16 tones and DDS up/down frequency conversion from DC--10\,GHz. They can generate pulses with arbitrary I-Q envelopes of variable length, modulating a carrier whose frequency, phase, and amplitude can be updated every fabric clock cycle (2.4\,ns) or as frequently as needed. The readouts support several types of filters and averages for two or more detection levels. The tProc runs the compiled quantum program that dispatches events to SGs, manages peripherals, and receives status and/or data from RDs. The tProc runs at the experiment's absolute clock and is able to execute multiple nested loops and conditional jumps for control and feed-forward. The XCOM allows multiple boards to be frequency- and phase-synchronized to an external reference, and allows multiple tProcs to synchronize to the same absolute clock. The absolute time is a 48-bit register associated with every tProc on every board. That absolute time can be started, stopped, or reset by the tProc, the software, or by an external input (I/O). The absolute clock wraps around after approximately 8 days, and does so without interrupting ongoing experiments. To synchronize PLLs, SGs, RDs, I/Os, and tProc absolute timing, we do the following:

\begin{itemize}[itemsep=2pt, parsep=0pt, topsep=4pt, partopsep=4pt]
    \item Connect the TI LMK04828B PLLs of all RFSoC boards to a stable external reference using matched-length cables.
    \item Configure and lock the LMK04828B on each board using the nested zero delay (ZDM) mode. The ZDM guarantees that all PLL outputs will have the same phase across all boards with sub-picosecond jitter.
    \item Synchronize all DAC blocks and ADC blocks on each board using the FPGA multi-tile sync (MTS) calibration. MTS uses PLL output clocks and SYSREF to synchronize DAC and ADC tiles. Each tile groups 4 DAC blocks or ADC blocks. For instance, MTS guarantees that DAC outputs will align to within 1 DAC sample time. For a 10\,GHz clock, this represents a maximum skew of 100\,ps on a single board and also across multiple boards.
\end{itemize}

Figure~\ref{fig:scope_RF} shows three RF waveforms from three different 
QICK boards and one PMOD digital I/O signal, all triggered simultaneously 
via matched-length cables. The PMOD drivers are bandwidth-limited, 
showing a 1.5\,ns rise time. As shown in the inset, the timing skew 
between the three RF channels is only 20\,ps, demonstrating sub-100\,ps 
cross-board synchronization. This result is achieved through the 
combination of PLL nested zero-delay mode lock on each board, multi-tile 
synchronization (MTS) across DAC and ADC tiles, and absolute clock 
alignment across all boards via XCOM.

\section{Operation and Validation}\label{sec:XCOM_tests}

The XCOM can be operated directly from the PS side (e.g., Python 
notebooks) or from the tProc, with its synchronization and communication 
functions embedded as part of the quantum algorithm. When operated 
from a Python script, some additional latency is introduced, intrinsic 
to an API running in a Linux environment; in that case the XCOM interface 
uses the FPGA AXI. When managed by the tProc, commands and data 
exchange run at the fabric clock (currently 430\,MHz), achieving the 
lowest latency.

The XCOM is managed by the tProc as a peripheral. Any algorithm in 
Python (or C++, etc.) can use the QICK tProc classes and functions, 
which are compiled into assembly machine language, loaded into the tProc 
program memory, and run independently at the absolute time clock 
frame~\cite{Stefanazzi2022}. The tProc XCOM classes, defined and 
integrated into the peripheral driver, provide commands to configure 
boards on the XCOM network, control the tProc state on other boards, 
and exchange data.

During network configuration, one board is designated the master and 
all others act as slaves. Since all XCOM interfaces and IPs are equal, 
any board can be assigned that role in software. A one-time XCOM 
power-up and synchronization sequence proceeds as follows, assuming 
every board is already synchronized to an external reference and, 
optionally, calibrated using MTS as described in 
Section~\ref{sec:XCOM_sync_sec}:

\begin{itemize}[itemsep=2pt, parsep=0pt, topsep=4pt, partopsep=4pt]
    \item Each board gets a network ID by running the \texttt{AUTO-ID} protocol. The protocol identifies the network port number and assigns it as the XCOM-ID for that board.
    \item The master board broadcasts a \texttt{RESET} absolute clock 
    command to synchronize.
    \item The master board broadcasts a \texttt{START} absolute clock 
    command. All tProcs begin running at the same time.
\end{itemize}

Once the network is running, any board can exchange data with any other 
board or broadcast to all boards. Data sizes are 8, 16, 32 bits, 
each preceded by a 4-bit board address field and a 4-bit command field. 
Communication latency is deterministic: 186\,ns per 32-bit word at the 
current prototype clock of $\sim$100\,MHz. Since the HP LVDS I/O and 
XCOM drivers support up to 312.9\,MHz, a factor of 3 faster, latency 
can be reduced to about 62\,ns per 32-bit word with a modification to the logic that runs the  
SERDES transceivers. The XCOM also provides a single flag bit, the 
fastest available message type, useful for point-to-point configurations 
such as a token ring network; this feature will be expanded in the future to be useful for a full mesh network.

The following validation tests have been performed on the XCOM prototype:

\begin{itemize}[itemsep=2pt, parsep=0pt, topsep=4pt, partopsep=4pt]
    \item \textbf{Synchronization stability:} The XCOM was synchronized 
    and configured once. Lock stability and tProc synchronization have 
    been tested to remain free of drift or loss of lock for multiple days.
    \item \textbf{Long-term deterministic latency:} 100K messages were 
    exchanged between 2 and 3 boards with identical latency over a 
    loopback path.
    \item \textbf{Conditional software jumps:} Wait for a message and 
    jump based on received data.
\end{itemize}

\section{Conclusion and Future Work}\label{sec:summary}

We have presented XCOM, a synchronization and low-latency communication 
network for QICK boards. The results shown are from a first 
prototype supporting up to five boards, operating at a clock speed below 
the maximum supported rate. The prototype successfully demonstrates the 
XCOM protocol in a real hardware environment, achieving a 32-bit word 
latency of 186\,ns that can be reduced to 62\,ns with a simple firmware 
modification. The prototype supports simultaneous exchange of multiple 
messages without central control, and sub-picosecond timing and phase 
synchronization have been demonstrated across three boards.

The current hardware design is not final; a more compact implementation 
is feasible for larger systems. XCOM enables control systems requiring 
accurate synchronization and communication, facilitating tasks such as 
auto-calibration of larger qubit systems with assistance from a central 
AI agent~\cite{li2026l,sivak2026}, as well as QEC syndrome detection 
with low-level decoders~\cite{Liyanage_2024,liyanage2025}.

In the future, the QICK team will continue developing deterministic, 
low-latency inter-board communication systems that scale without 
bottlenecks to high-speed control of large qubit systems. In the near term, we will port and expand XCOM to the AMD Versal RF platform~\cite{AMDversalrf}.

\bibliography{bibliography1.bib}

\end{document}